\documentclass[10pt, paper=a4, UKenglish]{article}
\usepackage{graphicx}
\usepackage[czech, english]{babel}
\usepackage[utf8]{inputenc}
\usepackage{todonotes}
\usepackage{algorithm}
\usepackage[noend]{algpseudocode}
\usepackage{amsmath}        %
\usepackage{amsfonts}       %
\usepackage{amsthm}         %
\usepackage{upgreek}

\newtheorem{defn}{Definition}

\newcommand{\CC}{C\nolinebreak\hspace{-.05em}\raisebox{.4ex}{\tiny\bf +}\nolinebreak\hspace{-.10em}\raisebox{.4ex}{\tiny\bf +}}
\def\CC{{C\nolinebreak[4]\hspace{-.05em}\raisebox{.4ex}{\tiny\bf ++}}}

\def\Title#1{\begin{center} {\Large #1 } \end{center}}
\def\Author#1{\begin{center}{ \sc #1} \end{center}}
\def\Address#1{\begin{center}{ \it #1} \end{center}}

\newcommand\pubblock{\rightline{\begin{tabular}{l} Proceedings of~the CTD/WIT 2019\\ \pubnumber\\
         \pubdate  \end{tabular}}}

\newenvironment{Abstract}{\begin{quotation} \begin{center} 
             \large ABSTRACT \end{center}\bigskip 
      \begin{center}\begin{large}}{\end{large}\end{center} \end{quotation}}

\newenvironment{Presented}{\begin{quotation} \begin{center} 
             PRESENTED AT\end{center}\bigskip 
      \begin{center}\begin{large}}{\end{large}\end{center} \end{quotation}}

\def\Acknowledgements{\bigskip  \bigskip \begin{center} \begin{large}
      \bf ACKNOWLEDGEMENTS \end{large}\end{center}}

\def\beq{\begin{equation}}
\def\eeq#1{\label{#1}\end{equation}}
\def\eeqn{\end{equation}}

\def\beqa{\begin{eqnarray}}
\def\eeqa#1{\label{#1}\end{eqnarray}}
\def\eeqan{\end{eqnarray}}

\let\bar=\overbar

\def\Dslash{\not{\hbox{\kern-4pt $D$}}}
\def\dslash{\not{\hbox{\kern-2pt $\del$}}}

\def\msb{{\bar{\ssstyle M \kern -1pt S}}}

\usepackage{color}
\usepackage{lineno}
\usepackage{subfig}
\usepackage{hyperref}

\newcommand\pubnumber{PROC-CTD19-105}

\newcommand\pubdate{\selectlanguage{english}\today}

\def\affiliation{
On behalf of\\
\textsuperscript{1} Institute of~Experimental and Applied Physics\\ 
Czech Technical University in Prague, Czech Republic\\
and\\
\textsuperscript{2} Faculty of Electrical Engineering\\
University of West Bohemia.
}

\newcommand{\conference}{Connecting the Dots and Workshop on Intelligent Trackers (CTD/WIT 2019)\\
Instituto de F\'isica Corpuscular (IFIC), Valencia, Spain\\ 
April 2-5, 2019}

\usepackage{fancyhdr}
\definecolor{mygrey}{RGB}{105,105,105}
\begin{document}

\large
\begin{titlepage}
\pubblock

\vfill
\Title{Real-time Timepix3 data clustering, visualization and classification\\ with a~new Clusterer framework}
\vfill

\Author{Lukáš Meduna\textsuperscript{1}, Benedikt Bergmann\textsuperscript{1}, Petr Burian\textsuperscript{1, 2},\\ 
Petr Mánek\textsuperscript{1}, Stanislav Pospíšil\textsuperscript{1}, Michal Suk\textsuperscript{1}
}
\Address{\affiliation}
\vfill

\begin{Abstract}
With the next-generation Timepix3 hybrid pixel detector, new possibilities and challenges have arisen. The Timepix3 segments active sensor area of~$1.98~cm^2$ into a~square matrix of~256 x 256 pixels. In each pixel, the Time of~Arrival (ToA, with a~time binning of~1.56 $ns$) and Time over Threshold (ToT, energy) are measured simultaneously in a data-driven, i.e. self-triggered, read-out scheme.

This contribution presents a~framework for data acquisition, real-time clustering, visualization, classification and data saving. All of~these tasks can be performed online, directly from multiple readouts through UDP protocol. Clusters are reconstructed on a~pixel-by-pixel decision from the stream of~not-necessarily chronologically sorted pixel data. To achieve quick spatial pixel-to-cluster matching, non-trivial data structures (quadtree) are utilized.

Furthermore, parallelism (i.e multi-threaded architecture) is used to further improve the performance of~the framework. Such real-time clustering offers the advantages of online filtering and classification of~events. Versatility of~the software is ensured by supporting all major operating systems (macOS, Windows and Linux) with both graphical and command-line interfaces.

The performance of~the real-time clustering and applied filtration methods are demonstrated using data from the Timepix3 network installed in the ATLAS and MoEDAL experiments at CERN.  
\end{Abstract}

\vfill

\begin{Presented}
\conference
\end{Presented}
\vfill
\end{titlepage}
\def\thefootnote{\fnsymbol{footnote}}
\setcounter{footnote}{0}
\normalsize

\section{Introduction}
\label{intro}

Timepix3 \cite{tpx3} is the latest addition to the family of~hybrid pixel detectors developed in the Medipix collaboration\footnote{https://medipix.web.cern.ch/}. 
Timepix3 features an active are of 1.98\,cm$^{2}$ segmented into a square matrix of 256 x 256 pixels at a pixel pitch of 55\,\,$\mu$m. Each pixel provides information about the energy deposition and the time of arrival (time binning 1.5625\,ns). Its high temporal granularity allows measurements in high flux environments without risking track overlap and was successfully employed to perform 3D reconstruction of particle tracks~\cite{BBsilicon, BBcdte}. Recently, Timepix3 detectors have been installed in the ATLAS~\cite{Burian_2018} and MoEDAL~\cite{MoEDAL} experiments at CERN, where they provide valuable information about the radiation levels and composition of the radiation fields.

Whereas previous chips of Timepix technology~\cite{timepix} purely rely on a frame-based readout scheme, Timepix3 comes with a data-driven operation, allowing a (quasi-)continuous measurement (per pixel dead-time: 475\,ns). In data driven mode, the output of the detector is a stream of individual pixel which can be processed almost in real-time. %
The aim of the presented work is to develop algorithms for fast track building to allow online visualization (for radiation level monitoring) and data filtering for pre-selection of relevant data to directly reject unwanted events and reduce disk space requirements. We have used the Katherine readout~\cite{kath} to test the developed methodology and algorithms. However, the implemented methods are fully compatible with other Timepix3 readouts~\cite{SPIDR, timewalk}.

\section{Clusterer framework}
\label{sec:clusterer-framework}

\subsection{Architecture}
The presented framework is written in~the modern \CC~14 language, which is suitable for high performance computations. The core of~the framework is independent of the GUI\footnote{GUI - graphical user interface.} that is written in~the Qt5 library\footnote{\url{https://doc.qt.io/qt-5/qt5-intro.html}}. Aside from the quadtree~\cite{quadtree} structure which bolsters the performance considerably, parallelism is used in~the framework. Individual steps (i.e.~reading input, parsing, clustering and visualizing) are separated into threads. Filtering of clusters is done using plug-ins compiled as dynamic libraries.

The framework was compiled, validated and tested on the major operating systems (Windows, macOS and Linux\footnote{Tested configurations -- Windows 10 64-bit edition, macOS 10.14 (Mojave) and Linux Mint 19.0 (Debian based distribution).}) thus ensuring the versatility of~the software. 
Both CLI\footnote{CLI - command line interface} and GUI versions were created.

\subsection{Event building}
\begin{defn}[Cluster of~pixels]\label{cluster-def}~\\
  Let $\Delta t$ be a~positive real number called size of~the time window, $X$ be a~set of~pixels and let for each pixel $p \in~X \;,\; ToA_p$ denotes its time of~arrival. A subset $C$ of~$X$ is called cluster, if for each pair of~pixels $p, p' \in~C$ the following two conditions are satisfied:
  \begin{enumerate}
      \item $|ToA_p - ToA_{p'}| < \Delta t \;$, i.e.~the times of~arrival of~the pixels $p$ and $p'$ differ less than $\Delta t$~\footnote{The value depends on the expected drift times of charge carriers and responses of pixel electronics. Conservatively, a value of 2000 $ns$ was chosen.}
      \item There exists a~path of~pixels in~$X$ from $p$ to $p'$, i.e.~a sequence of~8-connected pixels in~$X$ which starts in~$p$ and ends in~$p'$.
  \end{enumerate} 
\end{defn}

Individual cluster events are recreated based on spatial and time configuration using the \textit{ToA} and position in the sensor matrix (\textit{x,y}) of pixel hits. Due to the technical implementation in the Timepix3 chip, the last 200 $\mu$s of~pixels are not in chronological order.

\subsubsection{Pixel processing}
The goal is to transform partially unsorted hits from 2D space into the ToA sorted stream of separated events. The idea for processing incoming hits is given in~the following.

Keep a~set of~partial clusters (denoted as \textit{openClusters}). Try adding the new pixel to an existing cluster. If there is no existing cluster to which the new pixel can be added, create a~new cluster consisting just of~the current pixel. If the new pixel can be added to more than one cluster, join these clusters. This is achieved by Algorithm~\ref{alg:process}. 
If some of~the partial clusters contains only pixels with ToA less than $ToA - 200~\mu$s (the unordered time window) of~the new pixel, the cluster is outputted and removed from the set of \textit{openClusters}. Within these $200~\mu$s, hundreds of clusters can be under construction. 
\begin{algorithm} [H]
\caption{Process one pixel}\label{alg:process}
\begin{algorithmic}[1]
\Require
$openClusters \gets \left[ \right] $
\Procedure{ProcessPixel}{$timepixel$}\Comment{Pixel containing coordinates and ToA}
  \State $\mathit{added} \gets \mathit{False}$
  \For{\textbf{each}  $\mathit{cluster}$ \textbf{in} $ \mathit{openClusters}$}
      \If {$\Call{CanBeAdded}{cluster,timepixel}$}
          \If {$\mathit{added}$}
              \State $ lastCluster \gets \Call{JoinClusters}{cluster,lastCluster}$
          \Else
              \State $\Call{AddPixel}{cluster,timepixel}$
              \State $\mathit{added} \gets \mathit{True}$
              \State $\mathit{lastCluster} \gets \mathit{cluster}$
          \EndIf
      \EndIf
  \EndFor
  \If {\textbf{not} $added$}
      \State $\mathit{lastCluster} \gets \Call{CreateNewCluster}{openClusters}$
      \State $\Call{AddPixel}{lastCluster,timepixel}$
  \EndIf
  \State $\Call{CloseAndDispatchOldClusters}{openClusters}$
\EndProcedure
\end{algorithmic}
\end{algorithm}
The structure $timepixel$ is used to describe the tuple ($x$, $y$, ToA). The method \textit{CanBeAdded} uses internal quadtree structure to determine if the pixel is adjacent (i.e.~\textit{8-connected}) to the cluster. Joining clusters in~method \textit{JoinClusters} is done by transfer of~leaf nodes from the smaller tree to the bigger one. 

For quick validation, if a~pixel is adjacent to an open cluster, quadtree~\cite{quadtree} was selected and implemented. %

\subsubsection{Quadtree}
\label{subsec:quadtree}
For logarithmic access time and low memory consumption per cluster, the quadtree data structure was selected to represent the neighboring pixels during the construction of~separated events. The quadtree root node is representing a two-dimensional area. Each of~four child nodes corresponds to one of~the quadrants. At the end of~branches, the leaf nodes are the minimal area of~given size (called \textit{leaf size}). The complexity is $\mathcal{O}(\log_{4}n)$, where $n$ is number of~pixels in~the root area.

\begin{figure}[tbp]
  \centering
  \includegraphics[width=0.75\linewidth]{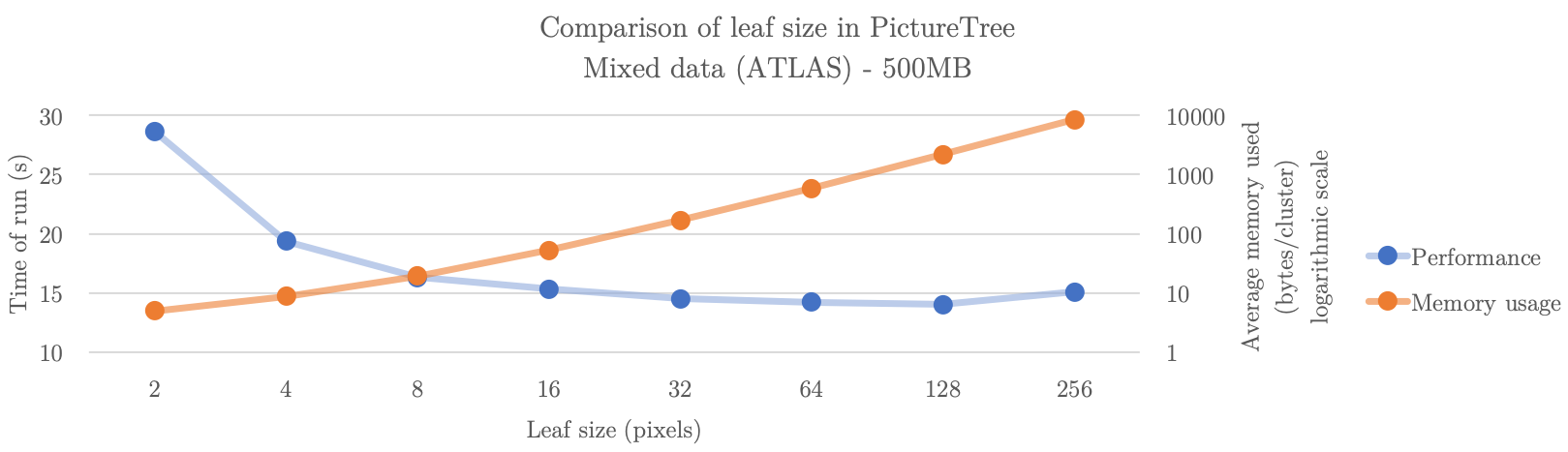}
  \caption{Performance test of different leaf sizes in~real-world data. The memory usage is compared with the time to process the dataset.}
  \label{fig:quadtree} 
\end{figure}

In the framework's implementation, an advanced approach was used. To minimize the memory allocations, the tree is built from bottom up. Thus, for a single pixel the tree size is equal to the \textit{leaf size} and it is increased when pixels outside the start area are added. To determine the optimal \textit{leaf size}, the performance was measured on real data from a~mixed radiation field in~ATLAS. The results are shown in Figure~\ref{fig:quadtree}. The size $16 \times 16$ was selected as a trade-off between performance and memory impact.

\section{Results}
\begin{figure}[tbp]
  \centering
  \hfill
  \subfloat[]{
     \includegraphics[width=0.45\linewidth]{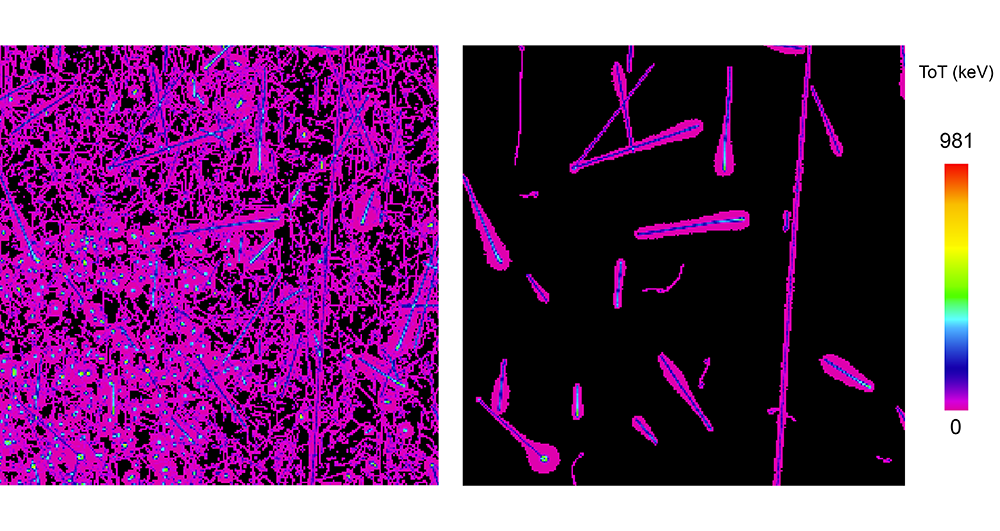}
  }
  \hfill
  \subfloat[]{
     \includegraphics[width=0.5\linewidth]{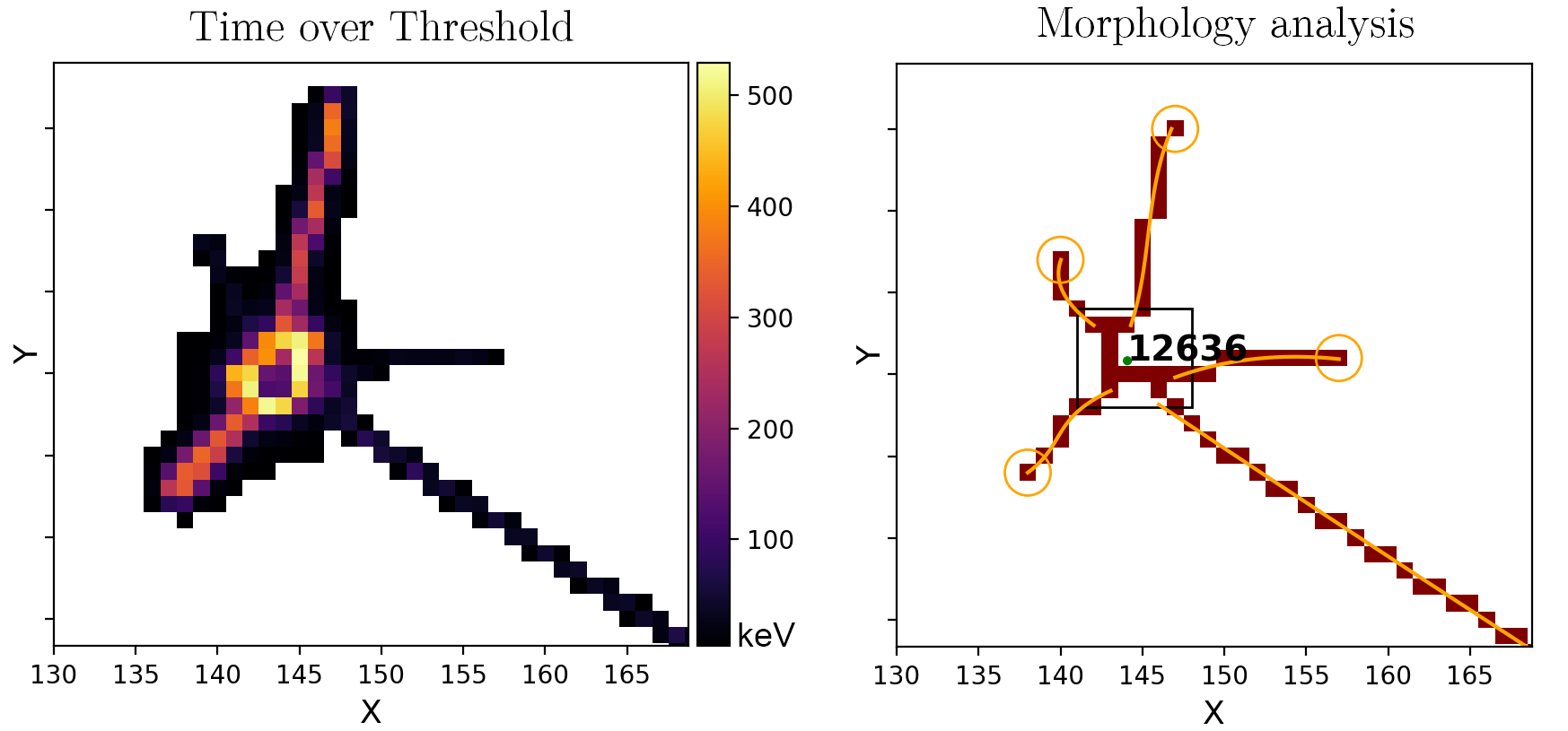}
  }
	\caption{(a) Integration window of 2 s without (left) and with (right) filtering of events of interest; (b) Single cluster representing an inelastic interaction with the silicon nucleus of the detection medium (fragmentation reaction), which was selected and further evaluated (skeletonized). Five outgoing tracks (prongs) could be identified. Data were acquired by Timepix3 installed in ATLAS.}
  \label{fig:complex}
\end{figure}

\begin{figure}[tbp]
  \centering
  \includegraphics[width=1\linewidth]{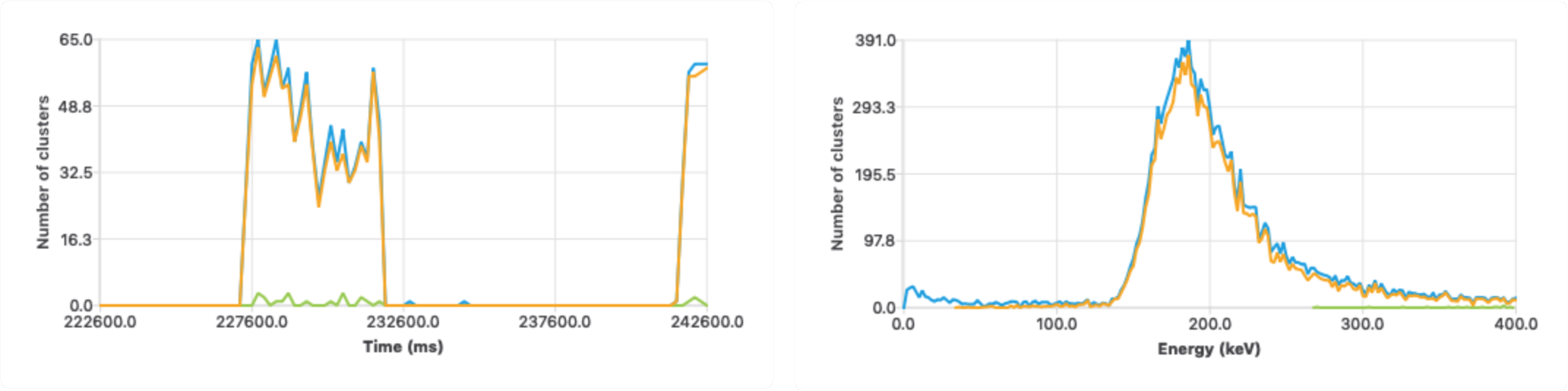}
  \caption{Real-time statistics of the measured radiation field. Cluster rate as a function of measurement time (left) and energy spectrum of the clusters during the measurement (right). The data were acquired during a test beam campaign in a relativistic particle beam at the Super-Proton-Synchrotron at CERN. The bottom image shows detail of the measurement while the filtering is applied. The blue line represents the raw input and the orange and green lines the filters based on class and size.}
  \label{fig:outputters}
\end{figure}

With the real-time event reconstruction, multiple metrics of~the radiation field can be observed in real-time. As an example, the cluster rates (radiation levels) and cluster energy spectrum as seen during a test beam measurement are given (Figure~\ref{fig:outputters}). The power (data reduction) of the real time data filtering is illustrated in Figure~\ref{fig:complex}(a), where all events within a 2\,s integration time are shown on the left and selected tracks of interest on the right. Proper filtering on track properties reduces the demand of disk space significantly without losing experiment relevant data. %
Pre-seleted Events of interest can then be further analyzed as shown in Figure~\ref{fig:complex}(b), where the tracks of fragmentation products are identified by skeletonization.

\section{Conclusion}
In this contribution, a comprehensive framework for data processing, event building and filtering on track features in the data-driven mode of Timepix3 was presented. Methods of~fast parallel and reliable clustering of~individual hits into separated events in~real-time were developed. The software can be used both offline (the source is a file) and online (the source is a Timepix3 device). The clusterer offers a variety of online measurement statistics (e.g. energy spectrum, event intensity and integration live display), with support of user-defined filtering.

The software contains support for energy calibration and time-walk correction techniques~\cite{timewalk,BBsilicon}. It can use the Katherine readout stream as direct data input. Modularity of the software allows interfacing different readouts or input sources. Online filtering on event properties was introduced as a powerful tool to reduce the amount of stored data for further analysis by regarding unwanted (background) events.

\Acknowledgements
The authors would like to express their sincere gratitude to the Medipix2 and Medipix3 Collaboration for their permanent support. We would like to thank to RNDr. František Mráz, CSc. of the Faculty of Mathematics and Physics, Charles University, for his advices during the development of the framework.

This work was supported by European Regional Development Funds: \textit{”Van de Graaf Accelerator and Tunable Source of~Monoenergetic Neutrons and Light Ions” (No. CZ.02.1.02 / 0.0 / 0.0 / 16 013 / 0001785)} and \textit{”Research Infrastructure for Experiments at CERN” (LM 2015058)}.

\end{document}